\documentclass[runningheads]{svmult}

\usepackage{makeidx}   
\usepackage{graphicx}  
\usepackage{subeqnar}  
\usepackage{multicol}  
\usepackage{physprbb}  
\makeindex             

\def\be{\begin{equation}}
\def\ee{\end{equation}}
\def\simless{\mathbin{\lower 3pt\hbox
   {$\rlap{\raise 5pt\hbox{$\char'074$}}\mathchar"7218$}}} 
\def\simgreat{\mathbin{\lower 3pt\hbox
   {$\rlap{\raise 5pt\hbox{$\char'076$}}\mathchar"7218$}}}
\def\kms{{\rm km\,s}^{-1}}

\begin{document}
\title*{Neutron Star Kicks and Asymmetric Supernovae}

\toctitle{Neutron Star Kicks}

\titlerunning{Neutron Star Kicks}

\author{Dong Lai}
\authorrunning{Dong Lai}

\institute{Center for Radiophysics and Space Research,
Department of Astronomy\\
Cornell University,
Ithaca, NY 14853, USA\\
Email: dong@astro.cornell.edu}

\maketitle              

\begin{abstract}
Observational advances over the last decade have left little doubt that 
neutron stars received a large kick velocity (of order a few hundred to a
thousand km\,s$^{-1}$) at birth. The physical origin of the kicks 
and the related supernova asymmetry is one of the central
unsolved mysteries of supernova research. We review the physics of 
different kick mechanisms, including hydrodynamically driven, 
neutrino -- magnetic field driven, and electromagnetically driven kicks. 
The viabilities of the different kick mechanisms are directly related to 
the other key parameters characterizing
nascent neutron stars, such as
the initial magnetic field and the initial spin. Recent
observational constraints on kick mechanisms are also discussed.
\end{abstract}

\section{Evidence for Neutron Star Kicks and Supernova Asymmetry}

It has long been recognized that neutron stars (NSs) have space velocities 
much greater (by about an order of magnitude) than their progenitors'. 
(e.g., Gunn \& Ostriker 1970). A natural explanation for such high velocities
is that supernova explosions are asymmetric, and provide kicks to the nascent
NSs.
In the last few years, evidence for NS kicks and supernova asymmetry 
has become much stronger. The observational facts and considerations 
that support (or even require) NS kicks fall into three categories:

\smallskip
\noindent
{\bf (1) Large NS Velocities ($\gg$ the progenitors' velocities 
$\sim 30$~km\,s$^{-1}$):} 

$\bullet$ Recent studies of pulsar proper motion give $200-500$~km\,s$^{-1}$ 
as the mean 3D velocity of NSs at birth (e.g., Lyne and Lorimer 1994; 
Lorimer et al.~1997; Hansen \& Phinney 1997; Cordes \& Chernoff 1998), 
with possibly a significant population having velocities greater than 
$1000$~km\,s$^{-1}$. While velocity of $\sim 100$~km\,s$^{-1}$ may in principle
come from binary breakup in a supernova (without kick), higher velocities 
would require exceedingly tight presupernova binary. Statistical analysis
seems to favor a bimodal pulsar velocity distribution, with peaks
around $100~\kms$ and $500~\kms$ (Arzoumanian et al.~2001; 
see also Hansen \& Phinney 1997; Cordes \& Chernoff 1998). 

$\bullet$ Direct evidence for pulsar velocities $\simgreat 1000$~km~s$^{-1}$ 
has come from observations of the bow shock produced by the Guitar Nebula 
pulsar (B2224+65) in the interstellar medium (Cordes, Romani \& Lundgren 1993).

$\bullet$ The studies of neutron star -- supernova remnant associations have, 
in many cases, indicated large NS velocities (e.g., Frail et 
al.~1994), although identifying the association can be
tricky sometimes (e.g. Kaspi 1999; Gaensler 2000). Of special interest is 
the recent studies of magnetar--SNR associations: the 
SGR 0526-66 - N49 association, implying $V_\perp\sim 
2900\,(3\,{\rm kyr}/t)$\,km\,s$^{-1}$, and the possible association 
of SGR 1900+14 with G42.8+0.6, implying $V_\perp\sim 1800\,
(10\,{\rm kyr}/t)$\,km\,s$^{-1}$. (However, the proper motion of
SGR 1806-20 may be as small as $100$~km~s$^{-1}$, and AXP 1E2259+586, 
AXJ 1845-0258, and AXP 1E1841-045 lie close to the centers of their 
respective remnants, CTB 109, G29.6+0.1, and Kes 73) (see Gaensler 2000).

\smallskip
\noindent
{\bf (2) Characteristics of NS Binaries (Individual Systems and 
Populations):} While large space velocities can in principle be accounted for
by binary break-up (as originally suggested by Gott et al.~1970; see Iben \&
Tutukov 1996), many observed characteristics of NS binaries 
demonstrate that binary break-up can not be solely responsible for 
pulsar velocities, and that kicks are required 
(see also Tauris \& van den Heuvel 2000). Examples include:

$\bullet$ The detection of geodetic precession in binary pulsar PSR 1913+16 
implies that the pulsar's spin is misaligned with the orbital angular momentum;
this can result from the aligned pulsar-He star progenitor only if
the explosion of the He star gave a kick to the NS that misalign 
the orbit (Cordes et al.~1990; Kramer 1998; 
Wex et al.~1999).

$\bullet$ The spin-orbit misalignment in PSR J0045-7319/B-star binary,
as manifested by the orbital plane precession 
(Kaspi et al.~1996; Lai et al.~1995) and fast orbital decay (which indicates
retrograde rotation of the B star with respect to the orbit; Lai 1996a;
Kumar \& Quataert 1997) require that the NS received a kick at birth (see
Lai 1996b).

$\bullet$ The observed system radial velocity ($430\,\kms$) of X-ray
binary Circinus X-1 requires $V_{\rm kick}\simgreat 500$\,km\,s$^{-1}$ 
(Tauris et al.~1999).

$\bullet$ High eccentricities of Be/X-ray binaries cannot be
explained without kicks (Verbunt \& van den Heuvel 1995).

$\bullet$ Evolutionary studies of NS binary population 
(in particular the double NS systems)
imply the existence of pulsar kicks (e.g., Deway \& Cordes 1987; Fryer \&
Kalogera 1997; Fryer et al.~1998). 

\smallskip
\noindent
{\bf (3) Observations of SNe and SNRs:}

There are many direct observations of nearby supernovae 
(e.g., spectropolarimetry: Wang et al.~2000, Leonard et al.~2000;
X-ray and gamma-ray observations and emission line profiles of SN1987A:
McCray 1993, Utrobin et al.~1995)
and supernova remnants 
(e.g., Morse, Winkler \& Kirshner 1995; Aschenbach et al.~1995) 
which support the notion that supernova explosions are not spherically
symmetric.

Finally it is of interest to note that recent study of 
the past association of the runaway star
$\zeta$ Oph with PSR J1932+1059 (Hoogerwerf et al.~2000) or with RX 185635-3754
(Walter 2000) also implies a kick to the NS. 

\section{The Problem of Core-Collapse Supernovae and Neutron Star Kicks}

The current paradigm for core-collapse supernovae leading to NS formation
is that these supernovae are neutrino-driven (see Bethe 1990,
Burrows 2000, Janka 2000 for recent review): As the central core of a
massive star collapses to nuclear density, it rebounds and sends off a
shock wave, leaving behind a proto-neutron star. The shock stalls
at several 100's km because of neutrino loss and
nuclear dissociation in the shock. A fraction of 
the neutrinos emitted from the proto-neutron star get absorbed
by nucleons behind the shock, thus reviving the shock, 
leading to an explosion on the timescale several 100's ms ---  
This is the so-called ``Delayed Mechanism''. However, 1D simulations with
detailed neutrino transport seem to indicate that neutrino heating of the
stalled shock, by itself, does not lead to an explosion or produce 
the observed supernova energetics (see Rampp \& Janka 2000). 
It has been argued that neutrino-driven convection in the proto-neutron star
(which tends to increase the neutrino flux) and that in the shocked mantle
(which tends to increase the neutrino heating efficiency) are central to the
explosion mechanism, although there is no consensus on the robustness
of these convections (e.g., Bethe 1990; Herant et al.~1994; 
Burrows et al.~1995; Janka \& M\"uller 1996; Mezzacappa et al.~1998).
What is even more uncertain is the role of rotation and magnetic field
on the explosion (see M\"onchmeyer et al.~1991; Rampp, M\"uller \& Ruffert 
1998; Khokhlov et al.~1999; Fryer \& Heger 2000 for simulations of 
collapse/explosion with rotation, and Thompson \& Duncan 1993 and
Thompson 2000a for discussion of possible dynamo processes and 
magnetic effects). 

It is clear that despite decades of theoretical investigations, our
understanding of the physical mechanisms of core-collapse supernovae remains 
significantly incomplete. The prevalence of neutron star kicks poses a
significant mystery, and indicates that large-scale, global deviation from
spherical symmetry is an important ingredient in our understanding of
core-collapse supernovae (see Burrows 2000).  

In the following sections, we review different classes of 
physical mechanisms for generating 
NS kicks (\S\S 3-5), and then discuss possible 
observational constraints and astrophysical implications (\S6). 
 
\section{Hydrodynamically Driven Kicks} 

The collapsed stellar core and its surrounding mantle are susceptible
to a variety of hydrodynamical (convective) instabilities 
(e.g., Herant et al.~1994; Burrows et al.~1995; Janka \& M\"uller 1996;
Keil et al.~1996; Mezzacappa et al.~1998). It is natural to expect that 
the asymmetries in the density, temperature and velocity distributions 
associated with the instabilities can lead to asymmetric matter ejection 
and/or asymmetric neutrino emission. Numerical simulations, however, indicate 
that the local, post-collapse instabilities are not adequate to account 
for kick velocities $\simgreat 100$~km~s$^{-1}$ (Janka \& M\"uller 1994;
Burrows \& Hayes 1996; Janka 1998; Keil 1998) --- These simulations were done
in 2D, and it is expected that the flow will be smoother on large scale in 
3D simulations, and the resulting kick velocity will be even smaller. 

There is now a consensus that global asymmetric perturbations 
in presupernova cores are required to produce the observed kicks
hydrodynamically (Goldreich, Lai \& Sahrling 1996; Burrows \& Hayes 1996).
Numerical simulations by Burrows \& Hayes (1996) demonstrate that if the
precollapse core is mildly asymmetric, the newly formed NS can
receive a kick velocity comparable to the observed values. (In one simulation,
the density of the collapsing core exterior to $0.9M_\odot$ and within $20^0$
of the pole is artificially reduced by $20\%$, and the resulting kick is about
$500$~km~s$^{-1}$.) Asymmetric motion of the exploding material (since the
shock tends to propagate more ``easily'' through the low-density region)
dominates the kick, although there is also contribution (about $10-20\%$)
from asymmetric neutrino emission. The magnitude of kick velocity is 
proportional to the degree of initial asymmetry in the imploding core.
Thus the important question is: What is the origin of the initial asymmetry? 

\subsection{Presupernova Perturbations}

Goldreich et al.~(1996) suggested that overstable g-mode oscillations in the
presupernova core may provide a natural seed for the initial asymmetry.
These overstable g-modes arise as follows. A few hours prior to core collapse, 
a massive star ($M\simgreat 8M_\odot$) has gone through a successive stages
of nuclear burning, and attained a configuration with a degenerate iron core
overlaid by an ``onion skin'' mantle of lighter elements. The rapidly growing
iron core is encased in and fed by shells of burning silicon and oxygen, and
the entire assemblage is surrounded by a thick convection zone. The nearly 
isothermal core is stably stratified and supports internal gravity
waves. These waves cannot propagate in the unstably stratified convection zone,
hence they are trapped and give rise to core g-modes in which the core
oscillates with respect to the outer parts of the star. The overstability 
of the g-mode is due to the ``$\varepsilon$-mechanism'' with driving provided
by temperature sensitive nuclear burning in Si and O shells surrounding 
the core before it implodes. It is simplest to see this
by considering a $l=1$ mode: If we perturb the core to the right, the 
right-hand-side of the shell will be compressed, resulting in an increase in
temperature; Since the shell nuclear burning rate depends sensitively
on temperature (power-law index $\sim 47$ for Si burning and $\sim 33$ for O
burning), the nuclear burning is greatly enhanced; this generates a large 
local pressure, pushing the core back to the left. 
The result is an oscillating g-mode with increasing amplitude. 

\begin{figure}
\begin{center}
\includegraphics[width=1.0\textwidth]{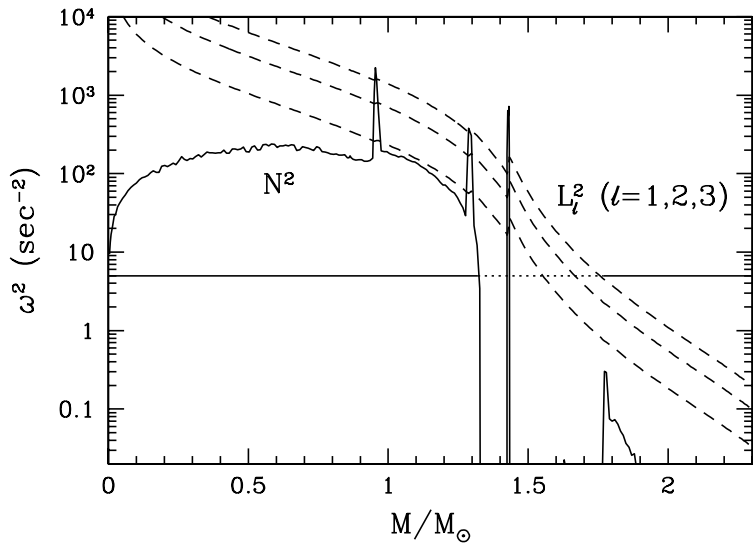}
\end{center}
\vskip -3truecm
\caption[]{Propagation diagram computed for a $15M_\odot$
presupernova model of Weaver and Woosley (1993). 
The solid curve shows $N^2$, where $N$ is the Brunt-V\"ais\"al\"a
frequency; the dashed curves show $L_l^2$, where $L_l$ is the 
acoustic cutoff frequency, with $l=1,~2,~3$.
The spikes in $N^2$ result from discontinuities in entropy and
composition. The iron core boundary is located at $1.3M_\odot$,
the mass-cut at $1.42M_\odot$.
Convective regions correspond to $N=0$. Gravity modes
(with mode frequency $\omega$) propagate in regions
where $\omega<N$ and $\omega<L_l$, while pressure modes
propagate in regions where $\omega>N$ and $\omega>L_l$.
Note that a g-mode trapped in the core can lose energy
by penetrating the evanescent zones and turning into an
outgoing acoustic wave (see the horizontal line). 
Also note that g-modes with higher $n$ (the
radial order) and $l$ (the angular degree) are better trapped in the
core than those with lower $n$ and $l$.}
\label{fig1}
\end{figure}

The main damping mechanism comes from the leakage of mode energy. 
The local (WKB) dispersion relation for nonradial waves is 
\be
k_r^2=(\omega^2c_s^2)^{-1}(\omega^2-L_l^2)(\omega^2-N^2),
\ee
where $k_r$ is the radial wavenumber, $L_l=\sqrt{l(l+1)}c_s/r$ ($c_s$ 
is the sound speed) and $N$ are the acoustic cut-off (Lamb) frequency and the 
Brunt-V\"ais\"al\"a frequency, respectively.
Since acoustic waves whose frequencies lie above the acoustic cutoff can
propagate through convective regions, each core g-mode will couple to an
outgoing acoustic wave, which drains energy from the core g-modes (see
Fig.~\ref{fig1}). This leakage of mode energy can be handled with an
outgoing propagation boundary condition in the mode calculation.
Also, neutrino cooling tends to damp the mode.
Since the nuclear energy generation rate depends more sensitively on
temperature than pair neutrino emission (power law index $\sim 9$), cooling is
never comparable to nuclear heating locally. Instead, thermal balance is
mediated by the convective transport of energy from the shells, where the
rate of nuclear energy generation exceeds that of neutrino energy
emission, to the cooler surroundings where the bulk of the neutrino
emission takes place. Calculations (based on the $15M_\odot$ and $25M_\odot$
presupernova models of Weaver \& Woosley 1993) indicate that 
a large number of g-modes are overstable, although for low-order modes
(small $l$ and $n$) the results depend sensitively on the detailed
structure and burning rates of the presupernova models.
The typical mode periods are $\simgreat 1$~s, the growth time
$\sim 10-50$~s, and the lifetime of the Si shell burning is $\sim$ hours 
(Lai \& Goldreich 2000b, in preparation).  

Our tentative conclusion is that overstable g-modes can potentially grow
to large amplitudes prior to core implosion, although a complete
understanding of the global pre-collapse asymmetries is probably out of reach
at present, given the various uncertainties in the presupernova 
models. For example, the O-Si burning shell is highly convective,
with convective speed reaching $1/4$ of the sound speed, and
hydrodynamical simulation may be needed to properly modeled such convection
zones (see Bazan \& Arnett 1998, Asida \& Arnett 2000). 
Alternatively, it has been suggested that the convection itself
may provide the seed of asymmetry in the presupernova core (Bazan \& Arnett
1998), although it is not clear whether the perturbations have sufficiently 
large scales to be relevant to supernova kicks. 


\subsection{Amplification of Perturbation During Core Collapse}

Core collapse proceeds in a self-similar fashion, with
the inner core shrinking subsonically and the outer core falling
supersonically at about half free-fall speed
(Goldreich \& Weber 1980; Yahil 1983). The inner
core is stable to non-radial perturbations because of the
significant role played by pressure in its subsonic collapse (Goldreich \&
Weber 1980). Pressure is less important in the outer region, making
it more susceptible to large scale instability.
A recent stability analysis of Yahil's self-similar collapse solution 
(which is based on Newtonian theory and a polytropic equation of state
$P\propto\rho^\Gamma$, with $\Gamma\sim 1.3$) does not reveal any unstable
global mode before the proto-neutron star forms (Hanawa \& Matsumoto 2000;
Lai 2000). However,  during the subsequent accretion of the outer core 
(involving $15\%$ of the core mass) and envelope onto the proto-neutron star,
nonspherical Lagrangian perturbations can grow according to
$\Delta\rho/\rho\propto
r^{-1/2}$ (independent of $l$) or even $\Delta\rho/\rho \propto r^{-1}$ 
(for $l=1$ when the central collapsed object is displaced from
the origin of the converging flow) (Lai \& Goldreich 2000) 
The asymmetric density perturbations seeded in the presupernova star,
especially those in the outer region of the iron core, are therefore amplified
(by a factor of 5-10) during collapse. The enhanced asymmetric density
perturbation may lead to asymmetric shock propagation and breakout, which then
give rise to asymmetry in the explosion and a kick velocity to the NS
(see Burrows \& Hayes 1996).  

\section{Neutrino -- Magnetic Field Driven Kicks}

The second class of kick mechanisms rely on asymmetric neutrino emission
induced by strong magnetic fields. 
Since $99\%$ of the NS binding energy (a few times $10^{53}$~erg) is
released in neutrinos, tapping the neutrino energy would appear to be an
efficient means to kick the newly-formed NS.
The fractional asymmetry $\alpha$ in the radiated neutrino energy required to
generate a kick velocity $V_{\rm kick}$ is $\alpha=MV_{\rm kick}c/E_{\rm tot}$
($=0.028$ for $V_{\rm kick}=1000$~km~s$^{-1}$, NS mass
$M=1.4\,M_\odot$ and total neutrino energy radiated $E_{\rm tot}
=3\times 10^{53}$~erg).

\subsection{Effect of Parity Violation}

Because weak interaction is parity violating, the neutrino opacities and
emissivities in a magnetized nuclear medium depend asymmetrically on the
directions of neutrino momenta with respect to the magnetic field, and 
this can give rise to asymmetric neutrino emission from the proto-neutron
star. Chugai (1984) (who gave an incorrect expression for the 
electron polarization in the relativistic, degenerate regime) 
and Vilenkin (1995) considered neutrino-electron scattering, but this is less
important than neutrino-nucleon scattering in determining neutrino transport in
proto-neutron stars. Dorofeev et al.~(1985) considered neutrino emission by
Urca processes, but failed to recognize that in the bulk interior of the star
the asymmetry in neutrino emission is cancelled by that associated with
neutrino absorption (Lai \& Qian 1998a). 

Horowitz \& Li (1998) suggested that large asymmetries in the neutrino flux
could result from the cumulative effect of multiple scatterings of neutrinos by
slightly polarized nucleons (see also Lai \& Qian 1998a; Janka 1998). However,
it can be shown that, although the scattering cross-section is asymmetric with
respect to the magnetic field for individual neutrinos, detailed balance
requires that there be no cumulative effect associated with multiple
scatterings in the bulk interior of the star where thermal equilibrium is
maintained to a good approximation (Arras \& Lai 1999a; see also Kusenko et
al.~1998).
For a given neutrino species, there is a drift flux of neutrinos along the
magnetic field in addition to the usual diffusive flux. This
drift flux depends on the deviation of the neutrino distribution function from
thermal equilibrium. Thus asymmetric neutrino flux can be generated 
in the outer region of the proto-neutron star (i.e., above the neutrino-matter
decoupling layer, but below the neutrinosphere) where the neutrino 
distribution deviates significantly from thermal equilibrium. While the
drift flux associated with $\nu_\mu$'s and $\nu_\tau$'s is exactly canceled by
that associated with $\bar\nu_\mu$'s and $\bar\nu_\tau$'s, there is a 
net drift flux due to $\nu_e$'s and $\bar\nu_e$'s. Arras \& Lai (1999b) 
found that the asymmetry parameter for the $\nu_e$-$\bar\nu_e$ flux is
dominated for low energy neutrinos ($\simless 
15$~MeV) by the effect of ground (Landau) state
electrons in the absorption opacity, $\epsilon_{\rm abs}\simeq
0.6B_{15}(E_\nu/1~{\rm MeV})^{-2}$, where 
$B_{15}=B/(10^{15}~{\rm G})$, and for high energy neutrinos by nucleon
polarization ($\epsilon\sim \mu_mB/T$). Averaging over all neutrino species, 
the total asymmetry in neutrino flux is of order
$\alpha\sim 0.2\epsilon_{\rm abs}$, and the resulting 
kick velocity $V_{\rm kick}\sim 50\,B_{15}$~km~s$^{-1}$.
There is probably a factor of 5 uncertainty in this estimate. To firm up
this estimate requires solving the neutrino transport equations 
in the presence of parity violation for realistic
proto-neutron stars.

\subsection{Effect of Asymmetric Field Topology}

A different kick mechanism relies on the asymmetric magnetic field 
distribution in proto-neutron stars (see Bisnovatyi-Kogan 1993;
however, he considered neutron decay, which is not directly relevant for
neutrino emission from proto-neutron stars). Since the cross section for
$\nu_e$ ($\bar\nu_e$) absorption on neutrons (protons) depends on the local
magnetic field strength due to the quantization of energy levels for the $e^-$
($e^+$) produced in the final state, the local neutrino fluxes emerged from
different regions of the stellar surface are different. Calculations
indicate that to generate a kick velocity of $\sim 300$~km~s$^{-1}$ using this 
mechanism alone would require that the difference 
in the field strengths at the two opposite poles of the star 
be at least $10^{16}$~G (Lai \& Qian 1998b). Note that unlike 
the kick due to parity violation (see \S 4.1), this mechanism does not
require the magnetic field to be ordered, i.e., only the magnitude
of the field matters. 

\subsection{Dynamical Effect of Magnetic Fields}

A superstrong magnetic field may also play a dynamical role in the
proto-neutron star. For example, it has been suggested that a locally strong
magnetic field can induce ``dark spots'' (where the neutrino flux is
lower than average) on the stellar surface by suppressing
neutrino-driven convection (Duncan \& Thompson 1992). While it is
difficult to quantify the kick velocity resulting from an asymmetric
distribution of dark spots, order-of-magnitude estimate indicates that a local
magnetic field of at least $10^{15}$~G is needed for this effect to be
of importance. Much work remains to be done to quantify 
the magnetic effects (especially when coupled with rotation)
on the dynamics of the proto-neutron star and the
supernova explosion (see, e.g., 
LeBlanc \& Wilson 1970; Thompson \& Duncan 1993).

\subsection{Exotic Neutrino Physics}

There have also been several ideas of pulsar kicks
which rely on nonstandard neutrino physics. It was
suggested (Kusenko \& Segre 1996) that asymmetric $\nu_\tau$ emission could 
result from the Mikheyev-Smirnov-Wolfenstein flavor transformation
between $\nu_\tau$ and $\nu_e$ inside a magnetized proto-neutron star
because a magnetic field changes the resonance condition for the
flavor transformation. This mechanism requires neutrino mass of order
$100$~eV. A similar idea (Akhmedov et al.~1997)
relies on both the neutrino mass and the neutrino magnetic 
moment to facilitate the flavor transformation (resonant neutrino
spin-flavor precession; see also Grasso et al.~1998;
Nardi \& Zuluaga 2000). More detailed 
analysis of neutrino transport (Janka \& Raffelt 1998), however, indicates
that even with favorable neutrino parameters (such as
mass and magnetic moment) for neutrino oscillation, the induced pulsar 
kick is much smaller than previously estimated (i.e.,
$B\gg 10^{15}$~G is required to obtain a $100$~km~s$^{-1}$ kick). 


\smallskip
It is clear that all the kick mechanisms discussed in this section (\S 4) 
are of relevance only for $B\simgreat 10^{15}$~G. While recent observations
have lent strong support that some neutron stars (``magnetars'')
are born with such a superstrong magnetic field (e.g., Thompson \& Duncan 1993;
Vasisht \& Gotthelf 1997; Kouveliotou et al.~1998,1999; Thompson 2000b), 
it is not clear (perhaps unlikely)
that ordinary radio pulsars (for which large velocities have been measured)
had initial magnetic fields of such magnitude (see als \S 6).

\section{Electromagnetically Drievn Kicks}

Harrison \& Tademaru (1975) show that electromagnetic (EM) radiation 
from an off-centered rotating magnetic dipole imparts
a kick to the pulsar along its spin axis. The kick is attained
on the initial spindown timescale of the pulsar (i.e.,
this really is a gradual acceleration), and
comes at the expense of the spin kinetic energy. We (Lai, Chernoff \& Cordes
2001) have reexamined this effect and found that the force on the pulsar 
due to asymmetric EM radiation is larger than the original 
Harrison \& Tademaru expression by a factor of four. If the dipole
is displaced by a distance $s$ from the rotation axis, and has components
$\mu_\rho,\mu_\phi,\mu_z$ (in cylindrical coordinates), the force
is given by (to leading order in $\Omega s/c$)
\be
F={8\over 15}\left({\Omega s\over c}\right){\Omega^4\mu_z\mu_\phi\over c^4}.
\ee
(The sign is such that negative $F$ implies ${\bf V}_{\rm kick}$
parallel to the spin $\bf\Omega$.) The dominant terms for the spindown 
luminosity give
\be
L={2\Omega^4\over 3c^3}\left(\mu_\rho^2+\mu_\phi^2+{2\Omega^2s^2\mu_z^2
\over 5c^2}\right).
\ee

For a ``typical'' situation, $\mu_\rho\sim\mu_\phi\sim\mu_z$, the
asymmetry parameter $\epsilon\equiv F/(L/c)$ is of order $0.4(\Omega s/c)$.
For a given $\Omega$, the maximum $\epsilon_{\rm max}=\sqrt{0.4}=0.63$ is 
achieved for $\mu_\rho/\mu_z=0$ and $\mu_\phi/\mu_z=\sqrt{0.4}\,
(\Omega s/c)$. From $M\dot V=\epsilon (L/c)=-\epsilon (I\Omega\dot\Omega)/c$,
we obtain the kick velocity
\be
V_{\rm kick}\simeq 260\,R_{10}^2\left({{\bar\epsilon}\over 0.1}\right)
\!\!\left({\nu_i\over 1\,{\rm kHz}}\right)^2\left[1-\left({\nu\over\nu_i}
\right)^2\right]{\rm km~s}^{-1},
\label{kick1}\ee
where $R=10R_{10}$~km is the neutron star radius,
$\nu_i$ is the initial spin frequency, $\nu$ is the current spin frequency
of the pulsar, and
$\bar\epsilon=(\Omega_i^2-\Omega^2)^{-1}\int\!\epsilon\,d\Omega^2$.
For the ``optimal'' condition, with $\mu_\rho=0$,
$\mu_\phi/\mu_z=\sqrt{0.4 }\,(\Omega_i s/c)$, and 
$\epsilon=\sqrt{0.4}\,
\left[2\Omega_i\Omega/(\Omega^2+\Omega_i^2)\right]$, we find
\be
V_{\rm kick}^{(\rm max)}\simeq
1400\,R_{10}^2\left({\nu_i\over 1\,{\rm kHz}}\right)^2
{\rm km~s}^{-1}.
\label{kickmax}\ee
Thus, if the NS was born rotating at $\nu_i\simgreat 1$~kHz, 
it is possible, in principle, to generate spin-aligned kick
of a few hundreds km~s$^{-1}$ or even 1000~km~s$^{-1}$.

Equations (\ref{kick1}) and (\ref{kickmax}) assume that the rotational energy
of the pulsar entirely goes to electromagnetic radiation. Recent
work has shown that a rapidly rotating ($\nu\simgreat 100$~Hz)
NS can potentially lose significant angular momentum to 
gravitational waves generated by unstable r-mode oscillations
(e.g., Andersson 1998; Lindblom, Owen \& Morsink 1998;
Owen et al.~1998; Andersson, Kokkotas \& Schutz 1999; Ho \& Lai 2000). 
If gravitational radiation carries away the rotational energy of the NS
faster than the EM radiation does, then the electromagnetic rocket effect
will be much diminished (Gravitational radiation can also carry away
linear momentum, but the effect for a NS is negligible).
In the linear regime, the r-mode amplitude $\alpha\sim \xi/R$ (where 
$\xi$ is the surface Lagrangian displacement; see the references cited above
for more precise definition of $\alpha$) grows due to gravitational radiation
reaction on a timescale $t_{\rm grow}\simeq 19\,(\nu/1~{\rm kHz})^{-6}$~s.
Starting from an initial amplitude $\alpha_i\ll 1$, the mode grows
to a saturation level $\alpha_{\rm sat}$ in time $t_{\rm grow}\ln(\alpha_{\rm
sat}/\alpha_i)$ during which very little rotational energy is lost. 
After saturation, the NS spins down due to gravitational radiation
on a timescale
\be
\tau_{\rm GR}=\left|{\nu\over\dot\nu}\right|_{\rm GR}\simeq
100\,\alpha_{\rm sat}^{-2}\left({\nu\over 1\,{\rm kHz}}\right)^{-6}\,{\rm s},
\ee
(Owen et al.~1998). By contrast, the spindown time due to EM radiation alone is
\be
\tau_{\rm EM}=\left|{\nu\over\dot\nu}\right|_{\rm EM}
\simeq 10^7\,B_{13}^{-2}\left({\nu\over 1\,{\rm
kHz}}\right)^{-2}\,{\rm s},
\label{taukick}\ee
where $B_{13}$ is the surface dipole magnetic field in units of $10^{13}$~G.
Including gravitational radiation, the kick velocity becomes
\be
V_{\rm kick}\simeq 260\,R_{10}^2\left({{\bar\epsilon}\over 0.1}\right)
\!\left({\nu_i\over 1\,{\rm kHz}}\right)^2
{1\over \beta}\ln\left[{1+\beta\over 1+\beta\,(\nu/\nu_i)^2}\right]
{\rm km~s}^{-1},
\label{kick2}\ee
where in the second equality we have replaced $\epsilon$ by
constant mean value $\bar\epsilon$, and $\beta$ is defined by 
\be
\beta\equiv \left({\tau_{\rm EM}\over\tau_{\rm GR}}\right)_i\simeq
\left({\alpha_{\rm sat}\over 10^{-2.5}}\right)^{2}\left({\nu_i\over 1\,{\rm
kHz}}\right)^{4}B_{13}^{-2}.
\ee
For $\beta\ll 1$, equation (\ref{kick2}) becomes eq.~(\ref{kick1}); for
$\beta\gg 1$, the kick is reduced by a factor $1/\beta$. 

Clearly, for the EM rocket to be viable as a kick mechanism
at all requires $\beta\simless 1$. The value of $\alpha_{\rm sat}$ is
unknown. Analogy with secularly unstable bar-mode in a Maclaurin spheroid 
implies that $\alpha_{\rm sat}\sim 1$ is possible (e.g., Lai \& Shapiro 1995).
It has been suggested that turbulent dissipation in the boundary layer near the
crust (if it exists early in the NS's history) may limit 
$\alpha_{\rm sat}$ to a small value of order $10^{-2}$-$10^{-3}$ (Wu, Matzner
\& Arras 2000). The theoretical situation is not clear at this point
(see Lindblom et al.~2000 for recent simulations of nonlinear r-modes).

\section{Astrophysical Constraints on Kick Mechanisms}

In \S\S3-5 we have focused on the {\it physics} of different kick mechanisms.
All these mechanisms still have intrinsic physics uncertainties and require
more theoretical work. For example:

(1) For the hydrodynamical driven kicks, one needs to
better understand the structure of pre-SN core in order to determine whether
overstable g-modes can grow to large amplitudes; more simulation would be 
useful to pin down the precise relationship between the magnitudes of
the initial asymmetry and the kick velocity; 

(2) For the neutrino--magnetic field driven kicks, more elaborate neutrino
transport calculation is necessary to determine (to within a factor of $2$)
the value of $B$ needed to generate (say) $V_{\rm kick}=300\,\kms$;
 
(3) For the electromagnetically driven kicks, the effect of gravitational 
radiation (especially the r-mode amplitude) needs to be better understood. 

\smallskip
We now discuss some of the astrophysical/observational 
constraints on the kick mechanisms. 

\subsection{Initial Magnetic Field of NS}

The neutrino-magnetic field driven kicks (\S 4) require
initial $B\simgreat 10^{15}$~G to be of interest. While magnetars may 
have such superstrong magnetic fields at birth (e.g., Thompson \& Duncan 1993;
Kouveliotou et al.~1998,1999; Thompson 2000b), the situation is not clear
for ordinary radio pulsars, whose currently measured magnetic fields
are of order $10^{12}$~G. It is difficult for an initial large-scale
$10^{15}$~G to decay (via Ohmic diffusion or ambipolar diffusion) to
the canonical $10^{12}$~G on the relevant timescale of $10^3-10^7$~years
(see Goldreich \& Reisenegger 1992). However, one cannot rule out the
possibility that in the proto-neutron star phase, a convection-initiated 
dynamo generates a ``transient'' superstrong magnetic field, lasting 
a few seconds, and then the field 
gets destroyed by ``anti-dynamo'' as the convection 
ceases. Obviously if we can be sure that this is not possible, then we
can discard the mechanisms discussed in \S 4.

\subsection{Initial Spin of NS}

To produce sufficient velocity, the electromagnetic rocket effect (\S 5)
requires the NS initial spin period $P_i$ to be less than $1-2$ ms.
It is widely thought that radio pulsars are not born
with such a rapid spin, but rather with a more modest
$P_i=0.02-0.5$~s (e.g., Lorimer et al.~1993). The strongest argument for 
this comes for the energetics of pulsar nebulae (particularly Crab). 
But this is not without uncertainties. For example, a recent analysis of the
energetics of the Crab Nebula suggests an initial spin period $\sim
3-5$ ms followed by fast spindown on a time scale of 30 yr (Atoyan 1999).
As for the Vela pulsar, the energetics of the remnant do not yield an
unambiguous constraint on the initial spin. Also, the recent discovery 
of the 16~ms X-ray pulsar (PSR J0537-6910) associated with the
Crab-like supernova remnant N157B (Marshall et al.~1998)
in the Large Magellanic Cloud implies that at least some NSs are 
born with spin periods in the millisecond range. So at this point it may 
be prudent to consider $P_i\sim 1$~ms as a possibility (see also \S 6.5).

\subsection{Natal vs. Post-Natal Kicks}

There is a qualitative difference between natal kicks 
(including the hydrodynamical driven and neutrino--magnetic 
field driven kicks) and post-natal kicks.
Because it is a slow process, the Harrison-Tademaru effect
may have difficulty in explaining some of the characteristics of
NS binaries (even if the physics issues discussed in \S 5 work out
to give a large $V_{\rm kick}$), such as the spin-orbit misalignment in PSR
J0045-7319 (Kaspi et al.~1996) and PSR 1913+16 (Kramer 1999) needed to
produce the observed precessions. For example,
in the case of PSR J0045-7319 -- B star
binary: if we assume that the orbital angular momentum of the 
presupernova binary is aligned with the spin of the B star, then the current
spin-orbit misalignment can only be explained by a fast
kick with $\tau_{\rm kick}$ less than the post-explosion orbital period 
$P_{\rm orb}$. Similarly, a slow kick (with $\tau_{\rm kick}\simgreat 
P_{\rm orb}$)
may be inconsistent with the NS binary populations (e.g.,
Dewey \& Cordes 1987; Fryer \& Kalogera 1997; Fryer et al.~1998).
However, note that $\tau_{\rm kick}\sim \tau_{\rm EM}$ 
for the Harrison-Tademaru effect depends on value of $B$ [see
eq.~(\ref{taukick})], thus can be made much smaller than $P_{\rm orb}$ 
(which typically ranges from hours to several months or a few years at most
for relevant binaries) if $B$ is large. 

\subsection{Correlations Between Velocity and Other Properties of NS ?}

Despite some earlier claims to the contrary, 
statistical studies of pulsar population have revealed no correlation between
$V_{\rm kick}$ and magnetic field strength, or correlation between
the kick direction and the spin axis (e.g., Lorimer et al.~1995; 
Cordes \& Chernoff 1998; Deshpande et al.~1999). Given the large 
systematic uncertainties, the statistical results, by themselves,
cannot reliably constrain any kick mechanism.
For example, the magnetic field strengths required for the
neutrino-driven mechanisms are $\simgreat 10^{15}$~G, much larger
than the currently inferred dipolar surface fields of typical radio pulsars;
there are large uncertainties in using the polarization angle to determine
the pulsar spin axis; differential galactic rotation is important for 
distant NSs and cannot be accounted for unless the distance is known 
accurately and the NS has not moved far from its birth location;  
several different mechanisms (including binary 
breakup) may contribute to the observed NS velocities (see
Lai, Chernoff \& Cordes 2001). 

Recent observations of the Vela pulsar and the surrounding compact X-ray nebula
with the Chandra X-ray Observatory reveal a two sided asymmetric jet
at a position angle coinciding with the position angle of the pulsar's
proper motion (Pavlov et al.~2000;
see http://chandra.harvard.edu/photo/cycle1/vela/ for image)
The symmetric morphology of the nebula
with respect to the jet direction strongly suggests that the jet is
along the pulsar's spin axis. Analysis of the polarization angle
of Vela's radio emission corroborates this interpretation. 
Similar evidence for spin-velocity alignment also exists
for the Crab pulsar. Thus, while statistical analysis 
of pulsar population neither support nor rule out any 
spin-kick correlation, at least for the Vela and Crab pulsars,
the proper motion and the spin axis appear to be aligned.
Interestingly, both Crab and Vela pulsars have relatively small
transverse velocities (of order $100\,\kms$).

\subsection{The Effect of Rotation and Spin-Kick Alignment?}

The apparent alignment between the spin axis and proper motion
for the Crab and Vela pulsar raises an interesting question: Under what 
conditions is the spin-kick alignment expected for different kick 
mechanisms? Let us look at the three classes of mechanisms discussed
in \S\S 3-5 (see Lai, Chernoff \& Cordes 2001).

(1) Electromagnetically Driven Kicks: The spin-kick 
slignment is naturally produced. (Again, note that 
$P_i\sim 1-2$~ms is required to generate sufficiently large $V_{\rm kick}$).

(2) Neutrino--Magnetic Field Driven Kicks: The kick is imparted to the NS 
near the neutrinosphere (at 10's of km) on the neutrino diffusion time, 
$\tau_{\rm kick}\sim 10$~seconds. As long as the initial spin period 
$P_i$ is much less than a few seconds, spin-kick alignment is naturally 
expected.

(3) Hydrodynamically Driven Kicks: 
The low-order g-modes trapped in the presupernova core ($M\simeq 1.4M_\odot,
\,R\simeq 1500$~km) have periods of 1-2 seconds, much shorter than
the rotation period of the core (unless
the core possesses a dynamically important angular momentum after collapse), 
thus the g-modes are not affected by 
rotation. Also, since the rotational speed of the core is typically less 
than the speed of convective eddies 
($\simeq$1000-2000~km~s$^{-1}$, about $20\%$ of the sound speed) 
in the burning shell surrounding the iron core, rotation should 
not significantly affect the shell convection either.
Thus the development of large-scale presupernova (dipolar)
asymmetry is not influenced by the core rotation. But even
though the primary thrust to the NS (upon core
collapse) does not depend on spin, the net kick will be
affected by rotational averaging if the asymmetry pattern
(near the shock breakout) rotates with the matter at a period shorter than
the kick timescale $\tau_{\rm kick}$. 
Here the situation is more complicated
because the primary kick to the NS is imparted at a large radius,
$r_{\rm shock}\simgreat 100$~km (since this the location of the stalled
shock). To obtain effective spin averaging, we require the rotation period
at $r_{\rm shock}$ to be shorter than $\tau_{\rm kick}\sim 100$~ms
(this $\tau_{\rm kick}$ is the shock travel time at speed 
of $10^4\,\kms$ across $\sim 1000$~km, the radius of the mass cut enclosing
$1.4M_\odot$). Assuming angular momentum conservation, this translates into 
the requirement that the final NS spin period $P_i\simless 1$~ms.
We thus conclude that if rotation is dynamically unimportant
for the core collapse and explosion (corresponding to
$P_s\gg 1$~ms), then rotational averaging is inefficient and 
the hydrodynamical mechanism does not produce spin-kick alignment.

The discussion above is based on the standard picture of
core-collapse supernovae, which is valid as long as rotation does {\it not}
play a dynamically important role (other than rotational averaging)
in the supernova. If, on the other hand, rotation is dynamically important, 
the basic collapse and explosion may be qualitatively different (e.g.,
core bounce may occur at subnuclear density, the explosion is
weaker and takes the form of two-sided jets; 
see M\"onchmeyer et al.~1991; Rampp, M\"uller \& Ruffert 1998; Khokhlov et
al.~1999; Fryer \& Heger 1999). 
The possibility of a kick in such 
systems has not been studied, but it is conceivable 
that an asymmetric dipolar perturbation may be coupled
to rotation, thus producing spin-kick alignment.

It has been suggested that the presupernova core has negligible 
angular momentum and the pulsar spin may be generated by off-centered kicks
when the NS forms (Spruit \& Phinney 1998). It is certainly true that even with
zero precollapse angular momentum, some rotation can be produced in the
proto-neutron star (Burrows et al.~1995 reported a rotation period of order a
second generated by stochastic torques in their 2D simulations of supernova
explosions), although $P_i\simless 30$~ms seems difficult to get. 
In this picture, the spin will generally be perpendicular to the velocity; 
aligned spin-kick may be possible if the kick is the result
of many small thrusts which are appropriately oriented (Spruit \& Phinney 1998)
--- this might apply if small-scale convection were responsible for
the kick. But as discussed in \S 3, numerical simulations indicate that such
convection alone does not produce kicks of sufficient amplitude.
Therefore, spin-kick alignment requires that 
the proto-neutron star have a ``primordial'' rotation 
(i.e., with angular momentum coming from the presupernova core).  

Clearly, {\it if} spin-kick alignment is a generic feature for all
NSs, it can provide strong constraints on the kick mechanisms, supernova 
explosion mechanisms, as well as initial conditions of NSs.

\bigskip
\medskip\noindent
{\bf Acknowledgement}

I thank my collaborators Phil Arras, David Chernoff, Jim Cordes,
Peter Goldreich, and Yong-Zhong Qian for their important contributions 
and insight. This work is supported by NASA grants NAG 5-8356 and NAG 5-8484 
and by a fellowship from the Alfred P. Sloan foundation. I thank 
the organizers for the invitation, and the European Center for Theoretical 
Physics for travel support to participate the ``Neutron Star Interiors''
Workshop (Trento, Italy, June-July 2000). This paper is also based on
a lecture given at the ITP conference on ``Spin and Magnetism in 
Neutron Stars'' (Santa Barbara, October 2000). I acknowledge ITP 
at UCSB (under NSF Grant PHY99-07949) for support while this
paper was being completed.



\begin{thebibliography}{8.}
\addcontentsline{toc}{section}{References}

\bibitem{Akhmedov97}
Akhmedov, E.~K., Lanza, A., \& Sciama, D.~W. 1997, Phys. Rev. D, 56, 6117

\bibitem{}
Andersson, N. 1998, ApJ, 502, 708

\bibitem{}
Andersson, N., Kokkotas, K.D., \& Schutz, B.F. 1999, ApJ, 510, 846

\bibitem{arras99a}
Arras, P., \& Lai, D. 1999a, ApJ, 519, 745 

\bibitem{arras99b}
Arras, P., \& Lai, D. 1999b, Phys. Rev. D60, 043001

\bibitem{}
Arzoumanian, Z., Chernoff, D.F., \& Cordes, J.M. 2001, in preparation.

\bibitem{}
Aschenbach, B., Egger, R., \& Trumper, J. 1995, Nature, 373, 587.

\bibitem{}
Asida, S.M., \& Arnett, D. 2000, ApJ, in press (astro-ph/0006451).

\bibitem{}
Atoyan, A. M. 1999, AAp, 346, L49

\bibitem{}
Bazan, G., \& Arnett, D. 1998, ApJ, 496, 316

\bibitem{Bisnovatyi-Kogan93}
Bisnovatyi-Kogan, G.~S. 1993, Astron. Astrophys. Trans., 3, 287

\bibitem{}
Burrows, A. 2000, Nature, 403, 727

\bibitem{Burrows95}
Burrows, A., Hayes, J., \& Fryxell, B.A. 1995, ApJ, 450, 830

\bibitem{}
Burrows, A., \& Hayes, J. 1996, Phys. Rev. Lett., 76, 352

\bibitem{Chugai84}
Chugai, N.N. 1984, Sov. Astron. Lett., 10, 87

\bibitem{Cordes93}
Cordes, J.M., Romani, R.W., \& Lundgren, S.C. 1993, Nature, 362, 133

\bibitem{Cordes90}
Cordes, J.M., Wasserman, I., \& Blaskiewicz, M. 1990, ApJ, 349, 546

\bibitem{}
Cordes, J.M., \& Chernoff, D.F. 1998, ApJ, 505, 315

\bibitem{}
Deshpande, A.A., Ramachandran, R., \& Radhakrishnan, V. 1999,
A\&A, 351, 195 

\bibitem{Deway87}
Deway, R.~J., \& Cordes, J.~M. 1987, ApJ, 321, 780

\bibitem{Dorofeev85}
Dorofeev, O.F., et al. 1985, Sov. Astron. Lett., 11, 123 

\bibitem{Frail94}
Duncan, R.C., \& Thompson, C. 1992, ApJ, 392, L9.

\bibitem{Frail94}
Frail, D.A., Goss, W.M., \& Whiteoak, J.B.Z. 1994, ApJ, 437, 781

\bibitem{Fryer98}
Fryer, C., Burrows, A., \& Benz, W. 1998, ApJ, 498, 333

\bibitem{}
Fryer, C.~L., \& Heger, A. 2000, 541, 1033

\bibitem{Fryer97}
Fryer, C., \& Kalogera, V. 1997, ApJ, 489, 244 (erratum:
ApJ, 499, 520)

\bibitem{Goldreich96}
Gaensler, B.M. 2000, in ``Pulsar Astronomy --- 2000 and Beyond'' (ASP
Conf. Proceedings) (astro-ph/9911190).

\bibitem{Goldreich96}
Goldreich, P., Lai, D., \& Sahrling, M. 1996, in
``Unsolved Problems in Astrophysics", ed. J.N. Bahcall and
J.P. Ostriker (Princeton Univ. Press)

\bibitem{}
Gott, J.R., Gunn, J.E., \& Ostriker, J.P. 1970, ApJ, 160, L91.

\bibitem{}
Grasso, D., Nunokawa, H., \& Valle, J.W.F. 1998, Phys. Rev. Lett., 81, 2412

\bibitem{}
Gunn, J.E., \& Ostriker, J.P. 1970, ApJ, 160, 979

\bibitem{}
Hanawa, T., \& Matsumoto, T. 2000, PASJ, in press.

\bibitem{}
Hansen, B.M.S., \& Phinney, E.S. 1997, MNRAS, 291, 569

\bibitem{}
Herant, M., et al. 1994, ApJ, 435, 339

\bibitem{}
Harrison, E.R., \& Tademaru, E. 1975, ApJ, 201, 447

\bibitem{}
Ho, W.C.G., \& Lai, D. 2000, ApJ, 543, 386

\bibitem{Horowitz97b}
Hoogerwerf, R., de Bruijne, J.H.J., \& de Zeeuw, P.T. 2000,
ApJ, in press (astro-ph/0007436)

\bibitem{Horowitz97b}
Horowitz, C.J., \& Li, G. 1998, Phys. Rev. Lett., 80, 3694 

\bibitem{}
Iben, I., \& Tutukov, A.~V. 1996, ApJ, 456, 738

\bibitem{Janka98}                 
Janka, H.-T. 1998, in ``Neutrino Astrophysics", ed. M. Altmann et al. 
(Garching: Tech. Univ. M\"unchen) (astro-ph/9801320)

\bibitem{Janka}
Janka, H.-T. 2000, submitted to A\&A (astro-ph/0008432)

\bibitem{}
Janka, H.-Th., \& M\"uller, E. 1994, A\&A, 290, 496

\bibitem{}
Janka, H.-T., \& M\"uller, E. 1996, A\&A, 306, 167

\bibitem{Janka98b}
Janka, H.-T., \& Raffelt, G.G. 1998, Phys. Rev. D59, 023005

\bibitem{}
Johnston, H.~M., Fender, R.~P., \& Wu, K. 1999, MNRAS, 308, 415.

\bibitem{Kaspi96}
Kaspi, V.M., et. al. 1996, Nature, 381, 583

\bibitem{kaspi}
Kaspi, V.M. 1999, 
in Pulsar Astronomy (ASP Conference Series), ed. M. Kramer, N. Wex and
R. Wielebinski (astro-ph/9912284)

\bibitem{}
Keil, W. 1998, in Proc. of the 4th Ringberg Workshop on Neutrino
Astrophysics (Munich, MPI).

\bibitem{Keil}
Keil, W., Janka, H.-Th., M\"uller, E. 1996, ApJ, 473, L111

\bibitem{}
Khokhlov, A.M., et al.~1999, ApJ, 524, L107 

\bibitem{}
Kouveliotou, C., et al. 1998, Nature, 393, 235

\bibitem{}
Kouveliotou, C., et al. 1999, ApJ, 510, L115

\bibitem{}
Kramer, M. 1998, ApJ, 509, 856

\bibitem{Kusenko96}
Kusenko, A., \& Segr\'e, G. 1996, Phys. Rev. Lett., 77, 4872 

\bibitem{Kusenko98a}                                       
Kusenko, A., Segr\'e, G., \& Vilenkin, A. astro-ph/9806205


\bibitem{}
Lai, D. 1996a, ApJ, 466, L35

\bibitem{}
Lai, D. 1996b, in Proceedings of the 18th Texas Meeting on Relativistic
Astrophysics, ed. A.V. Olinto et al. (World Scientific), p.~634
(astro-ph/9704134)

\bibitem{}
Lai, D. 2000, ApJ, 540, 946 

\bibitem{}
Lai, D., Bildsten, L., \& Kaspi, V.M. 1995, ApJ, 452, 819

\bibitem{}
Lai, D., Chernoff, D.F., \& Cordes, J.M. 2001, ApJ, in press
(astro-ph/0007272)

\bibitem{}
Lai, D., \& Goldreich, P. 2000a, ApJ, 535, 402

\bibitem{}
Lai, D., \& Goldreich, P. 2000b, in preparation

\bibitem{Lai98a}                         
Lai, D., \& Qian, Y.-Z. 1998a, ApJ, 495, L103 (erratum: 501, L155)

\bibitem{Lai98b}
Lai, D., \& Qian, Y.-Z. 1998b, ApJ, 505, 844

\bibitem{}
LeBlanc, J.M., \& Wilson, J.R. 1970, ApJ, 161, 541

\bibitem{}
Leonard, D.C., Filippenko, A.V., \& Ardila, D.R. 2000, ApJ, submitted
(astro-ph/0009285)

\bibitem{}
Lindblom, L., Owen, B.J., \& Morsink, S.M. 1998, Phys. Rev. Lett.,
80, 4843

\bibitem{}
Lindblom, L., Tohline, J.E., \& Vallisneri, M. 2000,  astro-ph/0010653

\bibitem{}
Lorimer, D.R., Bailes, M., Dewey, R.J., \& Harrison, P.A. 1993, 
MNRAS, 263, 403

\bibitem{Lorimer97}
Lorimer, D.R., Bailes, M., \& Harrison, P.A. 1997, MNRAS, 289, 592

\bibitem{Lyne94}
Lyne, A.G., \& Lorimer, D.R. 1994, Nature, 369, 127

\bibitem{}
Marshall, F.~E., et al.~1998, ApJ, 499, L179

\bibitem{}
McCray, R. 1993, ARA\&A, 31, 175

\bibitem{Mezza}
Mezzacappa, A., et al.~1998, ApJ, 495, 911.

\bibitem{}
M\"onchmeyer, R., Sch\"afer, G., M\"uller, E., Kates, R.E. 1991, A\&A, 246,
417

\bibitem{Morse95}
Morse, J.A., Winkler, P.F., \& Kirshner, R.P. 1995, AJ, 109, 2104

\bibitem{}
Owen, B.J., Lindblom, L., Cutler, C., Schutz, B.F.,
Vecchio, A., \& Andersson, N. 1998, Phys. Rev. D, 58, 084020

\bibitem{}
Pavlov, G.~G., et al. 2000, BAAS, 32, 733

\bibitem{}
Rampp, M., \& Janka, H.-Th 2000, ApJ, 539, L33

\bibitem{}
Rampp, M., M\"uller, E., \& Ruffert, M. 1998, A\&A, 332, 969


\bibitem{Spruit}
Spruit, H., \& Phinney, E.S. 1998, Nature, 393, 139

\bibitem{}
Tauris, T., et al. 1999, MNRAS, 310, 1165

\bibitem{}
Tauris, T., \& van den Heuvel, E.P.J. 2000,  astro-ph/0001015

\bibitem{}
Thompson, C. 2000a, ApJ, 534, 915

\bibitem{}
Thompson, C. 2000b, in Proceedings NATO Advanced Study Institute ``The
Neutron Star-Black Hole Connection'', ed. V. Connaughton et al.
(astro-ph/0010016)

\bibitem{}
Thompson, C., \& Duncan, R.C. 1993, ApJ, 408, 194.
 

\bibitem{Thompson96}
Thompson, C., \& Duncan, R.~C. 1996, ApJ, 473, 322

\bibitem{Utrobin95}
Utrobin, V.P., Chugai, N.N., \& Andronova, A.A. 1995, A \& A, 295, 129

\bibitem{}
van den Heuvel, E.P.J., \& van Paradijs, J. 1997, ApJ, 483, 399.

\bibitem{}
Vasisht, G., \& Gotthelf, E.V. 1997, ApJ, L129

\bibitem{Vilenkin95}
Verbunt, F., \& van den Heuvel, E.P.J. 1995, in X-ray Binaries, ed.
W.H.G. Lewin et al (Cambridge Univ. Press), p.457.

\bibitem{Vilenkin95}
Vilenkin, A. 1995, ApJ, 451, 700

\bibitem{wang}
Walter, F.M. 2000, ApJ, in press (astro-ph/0009031)

\bibitem{wang}
Wang, L., Howell, D.A., H\"oflich, P., \& Wheeler, J.C. 2000,
ApJ, submitted (astro-ph/9912033)

\bibitem{wex}
Weaver, T.A., \& Woosley, S.E. 1993, Phys. Rep., 227, 65

\bibitem{wex}
Wex, N., Kalogera, V., \& Kramer, M. 2000, ApJ, 528, 401 

\bibitem{}
Wu, Yanqin, Matzner, C.~D., \& Arras, P. 2000, ApJ, submitted
(astro-ph/0006123)


\end{thebibliography}
\end{document}